\documentclass[12pt]{article} 
\usepackage{amsmath}

\IfFileExists{amssymb.sty}
{ 
  \usepackage{amssymb}
  \newcommand{\rz}{{\mathbb{R}}} 
  \newcommand{\cz}{{\mathbb{C}}} 
  \newcommand{\nz}{{\mathbb{N}}} 
}
{ 
  \def\rz{\ifmmode {I\hskip -3pt R} \else {\hbox {$I\hskip -3pt R$}}\fi}
  \def\nz{\ifmmode {I\hskip -3pt N} \else {\hbox {$I\hskip -3pt N$}}\fi}
  \def\cz{\ifmmode {C\hskip -4.8pt\vrule height5.8pt\hskip 6..3pt} \else
  {\hbox {$C\hskip -4.8pt\vrule height5.8pt\hskip 6.3pt$}}\fi}
}

\IfFileExists{amsthm.sty}
{ 
  \usepackage{amsthm}
  \def\qed{\hbox {\hskip 1pt \vrule width 6pt height 6pt depth 1.5pt
        \hskip 1pt}}
}
{ 
\def\qed{\hbox {\hskip 1pt \vrule width 6pt height 6pt depth 1.5pt
        \hskip 1pt}}
\newenvironment{proof}{\textit{Proof.}}{\qed \flushright}
}

\newcommand{\defeq}{:=}
\newcommand{\e}{{\mathrm{e}}}
\newcommand{\cJ}{{\mathcal{J}}}
\newcommand{\cK}{{\mathcal{K}}}
\newcommand{\cL}{{\mathcal{L}}}
\def\tr{\mathop{\mathrm{tr}} \nolimits} 

\newtheorem{theorem}{Theorem}
\newtheorem{lemma}[theorem]{Lemma}

\newtheorem{corollary}[theorem]{Corollary}

\title{ A sharp bound for an eigenvalue moment of the one-dimensional  
    Schr\"{o}dinger operator
    \footnotetext{\copyright 1998 by the
    authors.  Reproduction of this article, in its entirety, by any
    means is permitted for non-commercial purposes}
    \footnotetext{Appeared in Adv.\ Theor.\ Math. Phys.\ \textbf{2}, 
              no.\ 4 (1998), 719- 731.}}

\author{Dirk Hundertmark$^{1}$\phantom{}\thanks{On leave of NWF I 
        - Mathematik, Universit\"at Regensburg, D-93040 Regensburg.} , 
        Elliott~H.~Lieb$^{1}$,\\ 
        and Lawrence E.~Thomas$^{2}$\\ 
        \footnotesize \it $^1$Departments of Physics and Mathematics, 
        Jadwin Hall, Princeton University,\\
        \footnotesize \it P.~O.~Box 708, Princeton, New Jersey 08544\\ 
        \footnotesize \it
        $^2$Department of Mathematics, University of Virginia,
        Charlottesville, Virginia 22903
}
\date{June 17, 1998 \\
      {\small (typos corrected Dec.\ 14, 1998)}} 

\begin{document}
\maketitle
\noindent
\begin{abstract}
  \noindent 
  We give a proof of the Lieb-Thirring inequality in the critical 
  case $d=1$, $\gamma= 1/2$, which yields the best possible constant.   
\end{abstract}

\section{Introduction}\label{section1} 
There is a family of inequalities \cite{LiebThirring1975}, 
\cite{LiebThirring1976} that has proved to be useful in various 
areas of mathematical physics, especially in the proofs of 
stability of matter. They state that given a Schr\"odinger 
operator 
\begin{displaymath}
    -\Delta + V \quad \mathrm{on~} \mathrm{L}^{2}(\rz^{d}), 
\end{displaymath}
the sum of the moments of the negative eigenvalues 
$-E_{1} < -E_{2} \leq -E_{3} \leq \ldots\leq 0$ (if any) of this 
operator is bounded by 
\begin{equation}
    \sum E_{i}^{\gamma} 
    \leq 
    L_{\gamma,d} \int (V_{-}(x))^{\gamma + d/2} \, dx, 
    \label{LiebThirringBound}
\end{equation}
with $V_{-}(x) \defeq \max(-V(x),0)$. 
These inequalities have been generalized in several directions, 
e.g. manifolds instead of $\rz^{d}$. Here we are concerned with 
the  case $d=1$.

The cases originally shown to hold \cite{LiebThirring1976} are
\begin{displaymath}
    d=1, \, \gamma > \frac{1}{2}, \quad d=2, \, \gamma > 0, \quad 
    \mathrm{and~} d\geq 3, \, \gamma \geq 0.
\end{displaymath}
When $d=2$ there cannot be any bound for $\gamma = 0$ (meaning the 
number of negative eigenvalues) since at least one negative 
eigenvalue always exists for arbitrarily small negative 
perturbations of the free Laplacian in two dimensions 
\cite[page 156-157]{LandauLifshitz}, \cite{Simon1976}.

The critical case $d \geq 3$ and $\gamma = 0$ was open for a while 
and proved independently by Cwikel \cite{Cwikel}, Lieb 
\cite{Lieb}, and Rozenbljum \cite{Rozenbljum}.
Still later, different proofs where given by Conlon \cite{Conlon} 
and Li and Yau \cite{LiYau}. The sharp constants are still not 
known, but the best one so far is in \cite{Lieb}.

If $d=1$ it is not hard to see that the inequality cannot hold 
for  $\gamma < 1/2$. To prove this choose a sequence of aproximate 
$\delta$-functions. They converge to zero in 
$\mathrm{L}^{\gamma+1/2}(\rz)$ but the limit may have a negative 
eigenvalue; see the discussion of a Dirac potential below.  
In the critical case $d=1, \gamma = 1/2$, which concerns us here, 
it was not known until recently whether $L_{1/2,1}$ is finite. 
This case was settled by Timo Weidl \cite{Weidl} who showed that 
$L_{1/2,1} < 1.005$. Unfortunately his method of proof 
\emph{cannot} be improved to yield the sharp constant as can be 
seen from the following argument: His method 
is also applicable for a half-line problem corresponding to a 
Schr\"odinger operator on $\rz_{+}$ with Neumann boundary 
conditions at the origin; in fact he reduces the full problem 
(but not the determination of the sharp constant) to this case. 
Since in this half-line problem the trivial lower bound for the 
sharp constant is given by $1$ his method cannot yield a better 
bound than $1$ in the problem concerning us here.

Hence, the sharp constant $L_{1/2,1}$ remained undetermined, a 
tantalizing situation, since there is an obvious conjecture 
about the value of this constant \cite{LiebThirring1976}. In 
one dimension the potential can be a measure (thanks to the fact 
that $H^{1}(\rz^{1})$ functions are continuous) and when 
$\gamma = 1/2$ the right hand side of (\ref{LiebThirringBound}) 
is simply the total mass of this measure.  
In order to maximize the sum of the square roots of the eigenvalues 
it is reasonable to suppose that one should concentrate the 
potential at one point and the extreme case should 
hence correspond to a $\delta$-function.  

It is well-known that $-\partial_{x}^{2} - c\delta$ is a well-defined 
closed quadratic form on the Sobolev space $H^{1}(\rz^{1})$ and the 
Hamiltonian corresponding to 
this form is used in textbooks as a simple solvable model in quantum 
mechanics. An exercise shows that the only bound state of this 
operator for positive $c$ is given by $\psi(x) = \exp(- c|x|/2)$ 
with eigenvalue $- c^{2}/4$.

If it is true that this Dirac potential is the optimal case we 
conclude that the sharp constant in the Lieb-Thirring inequality for 
$d=1, \gamma= 1/2$ is given by $L_{1/2,1}=1/2$.
The proof of this statement is the main result of this paper. A 
corollary of our result is that for the half-line problem with 
Neumann boundary conditions considered by Weidl, the sharp constant 
is $1$. 

Before turning to the proof let us note the corresponding -- still 
unproved -- conjecture when $1/2 < \gamma < 3/2$. The optimal 
potential should be given by
\begin{displaymath}
    V(x) =
    - \frac{1}{\gamma^{2} - 1/4} 
    \left(\cosh(\frac{x}{\gamma^{2} - 1/4}) \right)^{-2} 
\end{displaymath}
and the sharp constant is supposed to be \cite{LiebThirring1976}
\begin{displaymath}
    L_{\gamma,1} 
     = 
    \pi^{-1/2} \frac{1}{\gamma- 1/2} 
    \frac{\Gamma(\gamma +1)}{\Gamma(\gamma +1/2)}
    \left( \frac{\gamma -1/2}{\gamma +1/2} \right)^{\gamma +1/2}
     = 
    2L^{c}_{\gamma,1}
    \left( \frac{\gamma -1/2}{\gamma +1/2} \right)^{\gamma -1/2} .
\end{displaymath}
Here $L^{c}_{\gamma,1} \defeq (4\pi)^{-1/2} \Gamma(\gamma +1)/ 
\Gamma(\gamma+3/2)$ is its classical value.
Unlike the case $\gamma < 3/2$ the optimal constant in one 
dimension and $\gamma \geq 3/2$ is known \cite{AizenmanLieb}, 
\cite{LiebThirring1976} to be $L_{\gamma,1} = L^{c}_{\gamma,1}$.
Using the fact proved in \cite{AizenmanLieb} that 
$L_{\gamma,1} / L^{c}_{\gamma,1}$ is monotone decreasing in 
$\gamma$ and the sharp value for $L_{1/2,1}$ obtained here we 
conclude that $L_{\gamma,1} \leq 2 L^{c}_{\gamma,1}$ for all 
$\gamma\geq 1/2$. 
As a last remark, let us note that our proof uses no special 1-D 
technique, except for the explicit form of the Birman-Schwinger 
kernel (\ref{eq:BSkernel}) in one dimension.

\section{Proof of the main result for potentials}\label{section2}
The principal result of this paper is
\begin{theorem}\label{sharpinequality}
For a Schr\"odinger operator $-\partial_{x}^{2} + V$ in one 
dimension the optimal constant $L_{1/2,1}$ is $1/2$, 
i.e.\ 
\begin{equation}
    \sum_{-E_{i}\leq 0} \sqrt{\smash[b]{E_{i}}} 
    \,\leq \, 
    \frac{1}{2}\int V_{-}(x) \, dx
    \label{eq:sharpinequality} . 
\end{equation}
The inequality is strict if the negative part $V_{-}$ 
is a non-zero $\mathrm{L}^{1}$ function. 
\end{theorem}
In this section we prove this theorem in the case the potential 
is an $\mathrm{L}^{1}$ function. In the last section we extend 
the bound (\ref{eq:sharpinequality}) 
to potentials that are (finite) measures and prove that the 
$\delta$-function is the unique maximizer up to translations. By 
the minmax principle it suffices to investigate the operator 
$-\partial_{x}^{2} - V_{-} $.  We will henceforth assume 
$V=-U$ with $U$ non-negative and integrable.

To study the bound states energies of a Schr\"odinger operator 
it is often useful to investigate another problem. To do so we 
need some more notation. For $E > 0$ let  
\begin{equation}
    \cK_{E}(x,y) \defeq 
    \sqrt{\smash[b]{U(x)}}\, 
    \frac{\exp(-\sqrt{E}\,|x-y|)}{2\sqrt{E}} 
    \sqrt{\smash[b]{U(y)}}, 
    \quad \mathrm{for~all~} x,y \in \rz
    \label{eq:BSkernel}
\end{equation}
be the Birman-Schwinger kernel for the Schr\"odinger operator 
$-\partial_{x}^{2} - U$ in 
$\mathrm{L}^{2}(\rz)$. $\cK_{E}$ stands for the integral operator 
given by this kernel. The Birman-Schwinger principle 
\cite{Birman,Schwinger} states that $-E_{n} < 0$ is the 
$n^{\mathrm{th}}$ eigenvalue of $-\partial_{x}^{2} - U$ if and 
only if the $n^{\mathrm{th}}$ eigenvalue of $\cK_{E_{n}}$ equals 
one. The explicit expression (\ref{eq:BSkernel}) suggests that 
multiplying (\ref{eq:BSkernel}) by $\sqrt{E_{n}}$ will yield a still 
implicit but perhaps more flexible expression for $\sqrt{E_{n}}$. 
This is exactly what we are going to do. Let us define, for 
$\mu\geq 0$, 
\begin{equation}
    \cL_{\mu}(x,y) \defeq
    \sqrt{\smash[b]{U(x)}} 
     \e^{- \mu |x-y|} 
    \sqrt{\smash[b]{U(y)}}, 
    \quad \mathrm{for~all~} x,y \in \rz . 
    \label{eq:Lkernel}
\end{equation}
Moreover, given some arbitrary non-negative locally finite 
Borel-measure $\kappa$ on $\rz$, we can generalize the kernel 
(\ref{eq:Lkernel}) to 
\begin{equation}
    \cL^{\kappa}(x,y) \defeq
    \sqrt{\smash[b]{U(x)}} 
     \e^{-|\cJ(x)-\cJ(y)|} 
    \sqrt{\smash[b]{U(y)}}, 
    \quad \mathrm{for~all~} x,y \in \rz, 
    \label{eq:Lkappa}
\end{equation}
where the function $\cJ$ is given by 
\begin{equation}
    \cJ(x) \defeq
    \int_{0}^{x} \kappa(dz) .
    \label{eq:J}
\end{equation}
Again $\cL_{\mu}$ and $\cL^{\kappa}$ are the corresponding 
integral operators. Of course $\cL_{\mu}$ in (\ref{eq:Lkernel})  
corresponds to $\kappa(dz) = \mu dz$. 
Both $\cK_{E}$ and $\cL^{\kappa}$  are compact integral 
operators; their Hilbert-Schmidt norms are bounded by 
$(\int U(x) \, dx)^{2}/(2\sqrt{E})$ and $(\int U(x) \, dx)^{2}$, 
respectively. For a positive compact operator $A$ we denote its 
ordered eigenvalues by 
$\lambda_{1}(A)\geq\lambda_{2}(A)\geq\ldots\geq 0$. With the 
help of the Fourier transform   
($\exp(-\varepsilon|x|)/(2\varepsilon) = \int \e^{i px} / 
(p^{2}+\varepsilon^{2}) \,  dp /(2\pi)$)
one sees the following facts: \\[.3em]
\phantom{}~(i) $\cL^{\kappa}$ and $K_{E}$ are positive definite 
       operators, \\[.3em]
and hence the (ordered) eigenvalues $\lambda_{j}(\cL^{\kappa})$ 
obey \\[.3em]
\phantom{}~(ii) $\lambda_{1}(\cL^{\kappa}) > 
       \lambda_{2}(\cL^{\kappa}) \geq 
       \lambda_{3}(\cL^{\kappa}) \geq\ldots\geq 0$ \\[.3em]
with a similar statement for $\lambda_{j}(\cK_{E})$. The strict 
inequality follows from the positivity of the integral kernel and 
the Perron-Frobenius theorem.
The trace of $\cL^{\kappa}$ is given by \\[.3em]
\phantom{}~(iii) $\tr \cL^{\kappa} = \int U(x) \, dx$, \\[.3em]
independent of $\kappa$, and \\[.3em]
\phantom{}~(iv) $\cL^{0}= \cL_{0}$ is a rank one operator with 
    eigenvalue $\int U(x) \, dx$.\\[.3em]
The discussion above suggests that the sum of the square roots of 
the eigenvalues of the one dimensional Schr\"odinger operator is 
related to the sum of the eigenvalues of $\cL_{\mu}$. Indeed we 
have the following bound: 
\begin{theorem}[Domination by $\mathbf{\cL_{\mu}}$]\label{MainBound}
 Suppose $U\geq 0$ with $U\in \mathrm{L}^{1}(\rz)$ and let 
 $-E_{1} < -E_{2} \leq -E_{3} \leq \ldots\leq 0$  be the negative 
 eigenvalues counting multiplicity of the Schr\"odinger operator 
 $-\partial_{x}^{2} - U$ given by the minmax 
 principle. Furthermore, we denote by $\lambda_{j}(\cL_{\mu})$ 
 the eigenvalues of $\cL_{\mu}$ in (\ref{eq:Lkernel}). Then, for all 
 $n\in\nz$ and $0 \leq E \leq E_{n}$
 \begin{equation}
    2\sum_{i\leq n} \sqrt{\smash[b]{E_{i}}}
    \,\leq \,
    \sum_{i\leq n} \lambda_{i}(\cL_{\sqrt{\smash[b]{E}}}) + 
    \lambda_{1}(\cL_{\sqrt{\smash[b]{E_{1}}}}) 
    - \lambda_{1}(\cL_{\sqrt{\smash[b]{E_{2}}}}) .
    \label{mainbound}
  \end{equation}
  In (\ref{mainbound}) we set 
  $E_{j+1} = 0$ in case the Schr\"odinger operator happens to 
  have only $j$ negative eigenvalues. 
\end{theorem}
\begin{proof}
As already mentioned, the Birman-Schwinger principle gives a 
one-to-one correspondence between negative eigenvalues of a 
Schr\"odinger operator and the eigenvalues of $\cK_{E}$: 
$\lambda_{i}(\cK_{E_{i}}) = 1 $.
Multiplying this equality by $2\sqrt{E_{i}}$ yields 
$2\sqrt{\smash[b]{E_{i}}} =2\sqrt{\smash[b]{E_{i}}} \, 
\lambda_{i}(\cK_{E_{i}}) = 
\lambda_{i}(\cL_{\sqrt{\smash[b]{E_{i}}}}) $
for all $i$ such that $E_{i} >0$. Note that 
$\lambda_{i}(\cL_{0}) = 0$ if $i\geq 2$ since $\cL_{0}$ is 
a rank one operator. Therefore we have  
\begin{equation}
    2\sum_{i\leq n} \sqrt{\smash[b]{E_{i}}} 
    \, = \,
    \sum_{i\leq n} \lambda_{i}(\cL_{\sqrt{\smash[b]{E_{i}}}})
    \label{basicequality}
\end{equation}
for arbitrary  $n\in\nz$. If the eigenvalues of $\cL_{\mu}$ 
were monotonically decreasing as $\mu\geq 0$ increases this would 
immediately imply 
\begin{displaymath}
    2\sum_{i\leq n} \sqrt{\smash[b]{E_{i}}} 
    \leq 
    \sum_{i\leq n} \lambda_{i}(\cL_{\sqrt{\smash[b]{E_{n}}}})
    \leq
    \sum_{i\leq n} \lambda_{i}(\cL_{\sqrt{\smash[b]{E}}})
    \quad\mathrm{for~} 0\leq E\leq E_{n}.
\end{displaymath}
However, such a monotonicity \emph{cannot} hold since the trace 
of $\cL_{\mu}$ is \emph{independent} of $\mu\geq 0$. Nevertheless, 
\emph{the partial sums 
  $\sum_{i\leq n} \lambda_{i}(\cL_{\sqrt{\smash[b]{E}}})$ of its 
  eigenvalues are monotone in $E$} 
even for the slightly more general 
operator $\cL^{\kappa}$ given by (\ref{eq:Lkappa}). Lemma 
\ref{MonotonicityLemma} below is the key lemma in our analysis. 
Assuming the monotonicity given in Lemma \ref{MonotonicityLemma}, 
the proof of the theorem follows immediately from 
(\ref{basicequality}): For $n=1$ we have 
\begin{eqnarray*}
    2\sqrt{\smash[b]{E_{1}}} & = & 
    \lambda_{1}(\cL_{\sqrt{\smash[b]{E_{1}}}}) = 
    \lambda_{1}(\cL_{\sqrt{\smash[b]{E_{2}}}}) + 
    \lambda_{1}(\cL_{\sqrt{\smash[b]{E_{1}}}}) - 
    \lambda_{1}(\cL_{\sqrt{\smash[b]{E_{2}}}}) \\
     & \leq & 
    \lambda_{1}(\cL_{\sqrt{\smash[b]{E}}}) + 
    \lambda_{1}(\cL_{\sqrt{\smash[b]{E_{1}}}}) - 
    \lambda_{1}(\cL_{\sqrt{\smash[b]{E_{2}}}}) 
    \quad\mathrm{for~all~} 0\leq E\leq E_{1}
\end{eqnarray*}
where we take $E_{2} = 0$ if the potential has only one negative 
eigenvalue. If there are two or more negative eigenvalues it 
follows by induction that 
\begin{multline*}
    2\sum_{i\leq n} \sqrt{\smash[b]{E_{i}}} + 
    2\sqrt{\smash[b]{E_{n+1}}}  \\   
     \leq
    \sum_{i\leq n} \lambda_{i}(\cL_{\sqrt{\smash[b]{E_{n+1}}}}) + 
    \lambda_{n+1}(\cL_{\sqrt{\smash[b]{E_{n+1}}}})  
     + \lambda_{1}(\cL_{\sqrt{\smash[b]{E_{1}}}}) - 
    \lambda_{1}(\cL_{\sqrt{\smash[b]{E_{2}}}}) \\
     \leq  
    \sum_{i\leq n+1} \lambda_{i}(\cL_{\sqrt{\smash[b]{E}}})  + 
    \lambda_{1}(\cL_{\sqrt{\smash[b]{E_{1}}}}) - 
    \lambda_{1}(\cL_{\sqrt{\smash[b]{E_{2}}}}) 
\end{multline*}
for all $ 0\leq E\leq E_{n+1}$ and $n\in\nz$.
\end{proof}
Before proving the Lemma, we note a simple consequence of this 
theorem which proves our main bound (\ref{eq:sharpinequality}).
\begin{corollary}[Sharp constant]
Under the hypotheses of  Theorem \ref{MainBound} and for $U\neq 0$ 
\begin{displaymath}
    2\sum_{i\in\nz} \sqrt{\smash[b]{E_{i}}} 
    < 
    \int U(x) \, dx .
\end{displaymath}
\end{corollary}
\begin{proof}
{}From the theorem we get  
\begin{eqnarray}
    2\sum_{i\in \nz} \sqrt{\smash[b]{E_{i}}} 
     &\,\leq \, &
    \lambda_{1}(\cL_{0}) + 
    \lambda_{1}(\cL_{\sqrt{\smash[b]{E_{1}}}}) - 
    \lambda_{1}(\cL_{\sqrt{\smash[b]{E_{2}}}}) 
    \nonumber\\
     &=&
    \int U(x) \, dx + 
    \lambda_{1}(\cL_{\sqrt{\smash[b]{E_{1}}}}) - 
    \lambda_{1}(\cL_{\sqrt{\smash[b]{E_{2}}}}) , 
    \label{eq:finalbound}
\end{eqnarray}
since $ \cL_{0}$ is a rank one operator with eigenvalue  
$\int U(x) \, dx$. To conclude the strict inequality also note 
that $\lambda_{1}(\cL_{\sqrt{E}})$ 
is \emph{strictly} monotone decreasing in $E\geq 0$ by Lemma 
\ref{MonotonicityLemma}. The Perron-Frobenius theorem 
\cite[Theorem XIII.44]{ReedSimonIV} implies $E_{1}$ is simple 
and hence  $\lambda_{1}(\cL_{\sqrt{E_{1}}}) - 
\lambda_{1}(\cL_{\sqrt{E_{2}}}) < 0$. 
\end{proof}
\begin{lemma}[Monotonicity]\label{MonotonicityLemma}
 For all $n\in \nz$ the $n^{\mathrm{th}}$ partial sum of the 
 eigenvalues of the operator $\cL^{\kappa}$ defined in 
 (\ref{eq:Lkappa}) is monotonically decreasing in the sense 
 that 
 \begin{equation}
    \sum_{i\leq n} \lambda_{i}(\cL^{\kappa'})
    \leq
    \sum_{i\leq n} \lambda_{i}(\cL^{\kappa})
    \label{monotonicity}
 \end{equation}
 if $ \kappa'([s,t]) \geq \kappa([s,t])$ for all $s\leq t\in\rz$.
 Moreover the largest eigenvalue $\lambda_{1}(\cL^{\kappa})$ 
 is strictly monotone decreasing in $\kappa$.
\end{lemma}
\begin{proof}
 To clarify the line of reasoning we consider first a toy-model 
 given by an $(m\!+\!1)\!\times\! (m\!+\!1)$ matrix where the two 
 variables $x$ and $y$ in (\ref{eq:Lkappa}) take on $m+1$ values  
 $x_{0} \leq \ldots \leq x_{m}$. With 
 $a_{i} = \exp(-|J(x_{i}) - J(x_{0}|)) \leq 1$ (where $J$ is defined 
 in (\ref{eq:J}) and with $U=1$ 
 on $\{x_{0},\ldots,x_{m}\}$ for simplicity) the operator given in 
 (\ref{eq:Lkappa}) has the matrix
 \begin{displaymath}
    L(\!\{ a_{i}\}\!)
    \defeq
    \begin{pmatrix}
      1 & a_{1} & a_{1}a_{2} & a_{1}a_{2}a_{3} & \ldots & 
      a_{1} \ldots a_{m} \\
      a_{1} & 1 & a_{2}      & a_{2}a_{3}      & \ldots & 
      a_{2} \ldots a_{m} \\
      \vdots & & & & & \vdots \\
      a_{1}\ldots a_{m-1} & a_{2}\ldots a_{m-1} &\ldots & & 
      1 & a_{m} \\
      a_{1} \ldots a_{m} & a_{2} \ldots a_{m} & \ldots & & 
      a_{m} & 1 
    \end{pmatrix}.
 \end{displaymath}
 Let $\lambda_{1}(\!\{ a_{i}\}\!) \geq 
 \lambda_{2}(\!\{ a_{i}\}\!) \geq\ldots
 \geq\lambda_{m+1}(\!\{ a_{i}\}\!)$
 be the ordered eigenvalues of $    L(\!\{ a_{i}\}\!)$. 
 We investigate the sum of the largest $n$ eigenvalues in the 
 cube given by $0\!\leq\! a_{k} \leq 1$ for all 
 $k\in \{1,\ldots,m\!+\!1\}$ 
 and want to show that it is a (separately) monotone increasing 
 function of each $a_{k}$ in the interval $0\leq a_{k}\leq 1$. 
 Fix $k\in\{1,\ldots,m\!+\!1\}$ and $\{ a_{i}\}_{i\neq k}$. 
 For simplicity we write $L(a_{k})$ for 
 $L(\!\{ a_{i}\}_{i\neq k}, a_{k})$. The matrix $L$ has the form
 \begin{align*}
    L(a_{k}) \defeq L(\{a_{i}\}_{i\neq k}, a_{k}) &= 
    \begin{pmatrix}
      {A} & a_{k} W \\
       a_{k} W^{\dagger} & {B}
    \end{pmatrix}
    =:
    L(0) + a_{k} T \\
 \intertext{with} 
    L(0) \defeq L(\{a_{i}\}_{i\neq k}, 0) &= 
    \begin{pmatrix}
     {A} & \mathbf{0} \\
     \mathbf{0} & {B} 
    \end{pmatrix}
    \quad\mathrm{on~} \cz^{k}\oplus \cz^{m+1-k} = \cz^{m+1} \\
 \intertext{and the perturbation}
    T &=
    \begin{pmatrix}
     \mathbf{0} & {W} \\
     {W^{\dagger}} & \mathbf{0} 
    \end{pmatrix},
    \quad {W}:\, \cz^{n+1-k}\to \cz^{k},
 \end{align*}
  where ${A}$, ${B}$, and ${W}$ are $k\!\times\! k$, 
 $(m\!+\! 1\!-\!k)\!\times\! (m\!+\! 1\!-\!k)$, and 
 $k\!\times (m\!+\! 1\!-\!k)$ matrices respectively, depending 
 only on $\{a_{i}\}_{i\neq k}$. This shows that the dependence 
 of $L$ on $a_{k}$ (for fixed $\{a_{i}\}_{i\neq k}$) is affine-linear. 
 Now the claimed monotonicity of the sum of the largest $n$ eigenvalues 
 in $0\leq a_{k}\leq 1$ is easily seen by the usual quantum mechanics 
 textbook arguments of perturbation theory, cf. 
 \cite[chapter 3.5]{Thirring}: 
 The sum is given by
 \begin{eqnarray*}
    \sum_{i\leq n} \lambda_{i}(L(a_{k})) 
     &=&
    \sup_{0\leq d\leq 1,\; \tr d =n} 
    \tr(d L(a_{k}) ) \\
     &=&
    \sup_{0\leq d\leq 1,\;\tr d =n}
    \{\tr(d L(0) + a_{k} \tr(d T)\}
 \end{eqnarray*}
 where $d\! : \cz^{m+1} \rightarrow \cz^{m+1} $ is a density matrix. 
 Consequently, being a supremum of affine-linear functions, it is 
 convex. To conclude monotonicity in $a_{k}$ it is enough 
 to show that the derivative of the sum with respect to 
 $a_{k}$ at $a_{k}=0$ is non-negative. If the eigenvalues of 
 $L(0)$ are non-degenerate this follows immediately from the 
 Feynmann-Hellman theorem of perturbation theory: Since $L(0)$ leaves 
 the decomposition $\cz^{n+1}= \cz^{k} \oplus\cz^{m+1-k}$ invariant 
 its eigenvectors $\Phi_{i}$ live either in the subspace $\cz^{k}$  
 or $\cz^{m+1-k}$, so $\langle\Phi_{i},T\Phi_{i}\rangle = 0$. Thus by 
 the Feynman-Hellman formula each eigenvalue has derivative $0$ at 
 $a_{k} = 0$, and for this reason each partial sum has zero derivative 
 at $a_{k} =0$. 
 
 In the degenerate case a single eigenvalue might have 
 a negative derivative at $a_{k}=0$ but the partial sum of the 
 \emph{largest} $n$ eigenvalues always has a non-negative derivative. 
 Indeed, if the eigenvalues are degenerate we first have to diagonalize 
 the perturbation $T$ in the corresponding eigenspace $h$ of $L(0)$. 
 This eigenspace, however, can be decomposed into $h = h_{1} \oplus 
 h_{2}$, with $h_{1}\subset\cz^{k}$,  $h_{2}\subset\cz^{m+1-k}$, $h_{1}$ 
 or $h_{2}$ possibly empty. With $P_{i}$ being the orthogonal projection 
 onto $h_{i}$, $i= 1,2$, the perturbation $T$ restricted to the subspace 
 $h$ is again of the form $T|_{h} = P_{h} T P_{h} = \widetilde{W} + 
 \widetilde{W}^{\dagger}$, i.e., $ T |_{h} = 
   \left(\begin{smallmatrix}
    \mathbf{0} & \widetilde{W} \\
    \widetilde{W}^{\dagger} & \mathbf{0}
   \end{smallmatrix}\right) $ with  
 $\widetilde{W}\defeq P_{h_{1}} W P_{h_{2}} \!:  h_{2} \to h_{1} $.
 This gives $\tr_{h} T = \tr T|_{h}= 0$. The Feynman-Hellman formula 
 tells us that the eigenvalues of the restricted perturbation 
 $T|_{h}$ are the derivatives of the 
 eigenvalue branches emerging from this degeneracy subspace at 
 $a_{k}=0$. Since even the perturbation restricted to the 
 eigenspace $h$ has trace zero, we conclude that the derivative of 
 the sum at $a_{k}= 0$ is at most greater or equal to zero.
 
  For the strict monotonicity of the largest eigenvalue 
  $\lambda_{1}(L(\{a_{i}\}))$ in  the cube $0\!<\! a_{i}\!\leq\! 1$, 
  $i\in\{1,\ldots,m\!+\!1\}$ note that by the Frobenius--Perron 
  \mbox{theorem} 
  the corresponding eigenvector $\Phi(\{a_{i}\})$ has only positive 
  entries. Consequently for $0<a_{i}< a'_{i} \leq 1$, all 
  $i\in \{1,\ldots,m+1\}$, the minmax principle implies 
  \begin{eqnarray*}
    \lambda_{1}(L(\{a_{i}\})) 
     &=& 
    \langle
      \Phi(\{a_{i}\}),
        L(\{a_{i}\})
      \Phi(\{a_{i}\})
    \rangle \\
     &<&
    \langle\Phi(\{a_{i}\}),
      L(\{a'_{i}\})
    \Phi(\{a_{i}\})\rangle \\
     &\leq&
    \langle\Phi(\{a'_{i}\}),
      L(\{a'_{i}\})
    \Phi(\{a'_{i}\})\rangle
    =
    \lambda_{1}(\{a'_{i}\})
  \end{eqnarray*}

  \textbf{Remark:} The above reasoning for the toy model remains 
  valid if $L$ is replaced by $MLM$ where $M$ is a multiplication 
  operator, i.e.\ a diagonal matrix, so that the partial sums of the 
  eigenvalues for $MLM$ are also monotone.\\[0.3em]
  To apply this reasoning to our operator $\cL_{\mu}$ it is enough to 
  show the monotonicity (\ref{monotonicity}) for finite discrete 
  measures $\kappa = \sum c_{j} \delta_{x_{j}} $ and 
  $\kappa' = \sum c'_{j} \delta_{x_{j}} $ with $c'_{j}\geq c_{j}$.  
  Indeed, approximate $\kappa$ and $\kappa' - \kappa$ by finite 
  sums $\kappa_{m}$ and $\Delta_{m}$ 
  of $\delta$-functions. This is possible since they are weakly dense 
  in the set of locally finite Borel-measures. It is easy to see 
  that the corresponding operators $\cL^{\kappa_{m}}$ and 
  $\cL^{\kappa_{m} + \Delta_{m}}$ converge in Hilbert-Schmidt 
  norm to $\cL^{\kappa}$ and $\cL^{\kappa'}$. 
  Monotonicity of the partial sums of eigenvalues of 
  $\cL^{\kappa}$ for arbitrary $\kappa$ then follows by approximation 
  and, without loss of generality, we may assume 
  \begin{displaymath}
    \kappa = \sum_{j=1}^{m} c_{j} \delta_{x_{j}}, \quad
    \kappa' = \sum_{j=1}^{m} c'_{j} \delta_{x_{j}}
    \qquad\mathrm{~for~some~} m\in\nz
  \end{displaymath}
  with $c'_{j} \geq c_{j}\geq 0$, $j\in \{1,\ldots,m\}$, and 
  $-\infty < x_{1} < \ldots<x_{m} <\infty$. For $x\leq y$ we infer
  \begin{displaymath}
    |J(x) - J(y)| = \int_{x}^{ y} \kappa(dz) =
    \sum_{x \leq x_{j} \leq  y} c_{j}
  \end{displaymath}
  and
  \begin{eqnarray}
    \cL^{\kappa_{m}} (x,y) 
     & = & 
    \sqrt{U(x)} 
      \exp(-\!\!\sum_{x \leq x_{j} \leq y}\!\! c_{j})
    \sqrt{U(y)} \nonumber \\
     &=&
    \prod_{x \leq x_{j} \leq y}
     \e^{-c_{j}} \sqrt{U(x)}\sqrt{U(y)}
    \nonumber  \\
     & = & 
    \prod_{x \leq x_{j} \leq y}
     a_{j} \sqrt{U(x)}\sqrt{U(y)},
     \qquad a_{j} \defeq \e^{-c_{j}}, \, j=1,\ldots,m
     \nonumber \\
      & =: &
     \cL(\{a_{j}\}) (x,y) .
    \label{Discretekernel} 
  \end{eqnarray}
  As in the matrix case the dependence of $\cL(\{a_{i}\})$ on a 
  single $a_{k}$ (for fixed $\{a_{j}\}_{j\neq k}$) is affine-linear 
  and decomposition of the Hilbert space is now given by 
  $\mathrm{L}^{2}(\rz) =  \mathrm{L}^{2}(-\infty,x_{k}) \oplus 
   \mathrm{L}^{2}(x_{k},\infty)$.
  Hence we are in precisely the same situation as for our 
  $MLM$ toy--model, and we infer that the partial sums of 
  the largest eigenvalues are monotone in $\kappa$ for 
  $\cL^{\kappa_{m}}$. By the above limiting argument therefor 
  for $\cL^{\kappa}$ and in particular for $\cL_{\mu}$.
  
  Strict monotonicity of the largest eigenvalue 
  $\lambda_{1}(\cL^{\kappa})$ in $\kappa$, i.e.\ 
  $\lambda_{1}(\cL^{\kappa'}) < \lambda_{1}(\cL^{\kappa})$ if 
  $\kappa'>\kappa$, follows from the Perron-Frobenius 
  theorem, the minmax principle, and the strict monotonicity of the 
  kernel (\ref{eq:Lkappa}) in $\kappa$. One can, however, avoid the 
  minmax principle in this conclusion. The Perron-Frobenius theorem 
  states that the eigenvectors $\Phi_{1}^{\kappa}$ and 
  $\Phi_{1}^{\kappa'}$ corresponding to $\lambda_{1}(\cL^{\kappa})$ 
  and  $\lambda_{1}(\cL^{\kappa'})$ are non-negative and strictly 
  positive on the support of the potential $U$. By definition
  \begin{displaymath}
    \lambda_{1}(\cL^{\kappa})\Phi_{1}^{\kappa} 
    = \cL^{\kappa}\, \Phi_{1}^{\kappa}
  \end{displaymath}
  and the same for $\kappa'$. {}From this we get
  \begin{equation}
    \lambda_{1}(\cL^{\kappa'}) 
    \langle\Phi_{1}^{\kappa},\Phi_{1}^{\kappa'}\rangle -
    \lambda_{1}(\cL^{\kappa}) 
    \langle\Phi_{1}^{\kappa'},\Phi_{1}^{\kappa}\rangle
     =  
    \langle\Phi_{1}^{\kappa},\cL^{\kappa'}\Phi_{1}^{\kappa'}\rangle
    -
    \langle\Phi_{1}^{\kappa'},\cL^{\kappa}\Phi_{1}^{\kappa}\rangle.
    \label{direct1}
  \end{equation}
  since $\langle\Phi_{1}^{\kappa},\Phi_{1}^{\kappa'}\rangle > 0$ and 
  the scalar products in (\ref{direct1}) are real, hence symmetric, 
  we get by interchanging the integration variables 
  \begin{eqnarray*}
    \lambda_{1}(\cL^{\kappa'}) - \lambda_{1}(\cL^{\kappa}) 
     & = & 
     \frac{1}{\langle\Phi_{1}^{\kappa},\Phi_{1}^{\kappa'}\rangle}
     \iint \Phi_{1}^{\kappa}(x)\Phi_{1}^{\kappa'}(y)
     (\cL^{\kappa'}(x,y) - \cL^{\kappa}(x,y)) \, dxdy
     \\
     & < & 0
  \end{eqnarray*} 
  by the strict monotonicity of the kernel $\cL^{\kappa}(x,y)$ in 
  $\kappa$ and the strict positivity of $\Phi_{1}^{\kappa}$, 
  $\Phi_{1}^{\kappa'}$ on the support of $U$. This concludes the 
  proof of the monotonicity lemma.
\end{proof}

\section{Extension to `potentials' that are measures}\label{section3}
In this section we extend theorem \ref{sharpinequality} to 
measure perturbations of $-\partial_{x}^{2}$. As mentioned in the 
introduction the Sobolev inequality in one dimension, cf. 
\cite{LiebLoss}[Theorem 8.5], ensures that a finite measure $\tau$ 
on $\rz$ yields a quadratic form $\tau[\phi]\defeq \int |\phi(x)|^{2} \, 
\tau(dx)$ that is infinitesimally form bounded with respect to the 
Laplacian in one dimension. The quadratic form
\begin{eqnarray}
    \langle\psi, H\phi\rangle & = & 
    \langle\psi,-\partial_{x}^{2}\phi\rangle +
     \langle\psi,\tau\phi\rangle  
     \nonumber\\
    & \defeq & 
    \langle\partial_{x}\psi,\partial_{x}\phi\rangle
    +
    \int_{\rz} \overline{\psi(x)}\phi(x)\,\tau(dx)
    \label{quadraticform}
\end{eqnarray}
is thus closed on the Sobolev space $H^{1}(\rz)$ and defines   
a unique self-adjoint operator $H=-\partial_{x}^{2} +\tau$ on 
$\mathrm{L}^{2}(\rz)$. By the minmax 
principle for forms it is again enough to consider the case 
$\tau = - \nu$ for some positive bounded measure $\nu$ on $\rz$. We 
will hence consider $H=-\partial_{x}^{2} -\nu$. Our result is
\begin{theorem}
 Suppose $\nu$ is a non-negative measure with $\nu(\rz) < \infty$ 
 and let $-E_{1} < -E_{2} \leq -E_{3} \leq \ldots\leq 0$  be the 
 negative eigenvalues counting multiplicity of the Schr\"odinger 
 operator $-\partial_{x}^{2} - \nu$ (if any) given by the 
 corresponding quadratic form. Then  
 \begin{equation}
    \sum_{i=1}^{\infty} \sqrt{\smash[b]{E_{i}}}
    \,\leq \,
    \frac{1}{2}\, \nu(\rz)
    \label{eq:measurebound}
  \end{equation}
  with equality if and only if the measure $\nu$ is a 
  single Dirac measure. 
\end{theorem}
\begin{proof}
One obstacle in the proof of this theorem is to construct an analog 
of the Birman-Schwinger kernel (\ref{eq:BSkernel}) for measures. 
It is given by
\begin{equation}
    \widetilde{K}_{E}[\nu](x,y) \defeq
    \int \frac{1}{\sqrt{p^{2} +E}}(x,\zeta) 
    \frac{1}{\sqrt{p^{2} +E}}(\zeta,y)\,\nu(d\zeta)
\end{equation}
where we set $p^{2} \defeq -\partial_{x}^{2}$ for convenience. 
A given measure $\nu$ can be approximated by smooth functions by 
convoluting it with an approximate $\delta$-function $\nu\rightarrow 
\nu_{\varepsilon} = \delta_{\varepsilon} \ast \nu$. Of course, $ 
\nu_{\varepsilon} \to \nu$ weakly and the operators 
$\widetilde{K}_{E}[\nu_{\varepsilon}]$ converge to 
$\widetilde{K}_{E}[\nu]$ for large $E$ in Hilbert--Schmidt norm, 
hence in the usual operator norm, too. By Tiktopoulos' formula 
\cite{Simon1971} this shows the norm convergence of the resolvents 
$(p^{2} \!-\! \nu_{\varepsilon} \!+\! E)^{-1}$ to 
$(p^{2} \!-\! \nu \!+\! E)^{-1}$ and thus any finite collection of 
eigenvalues of $p^{2} -\nu_{\varepsilon}$ converges to those of 
$p^{2} -\nu$. So, applying the results of the last section, we have 
for any partial sum, i.e.\ any $n\in\nz$
\begin{eqnarray*}
	2\sum_{i\leq n} \sqrt{\smash[b]{E_{i}}}
	 &\leq&
	\lim_{\varepsilon \to 0} \int \nu_{\varepsilon}(x)\, dx + 
	\lim_{\varepsilon \to 0} \big(
	  \lambda_{1}(\cL_{\sqrt{E_{1}}}[\nu_{\varepsilon}]) - 
	  \lambda_{1}(\cL_{\sqrt{E_{2}}}[\nu_{\varepsilon}])
	\big)
	\\
	&=& \int \nu(dx) + 
	\lim_{\varepsilon \to 0} \big(
	  \lambda_{1}(\cL_{\sqrt{E_{1}}}[\nu_{\varepsilon}]) - 
	  \lambda_{1}(\cL_{\sqrt{E_{2}}}[\nu_{\varepsilon}])
	\big)
\end{eqnarray*}
where for $\mu \geq 0$ the operator $\cL_{\mu}[\nu_{\varepsilon}]$ 
is defined by the right hand side of (\ref{eq:Lkernel}) with $U(x)$ 
replaced by $\nu_{\varepsilon}(x)$. For any positive bounded measure 
$\nu$ let 
$\widetilde{\cL}_{\mu}[\nu] = 2\mu (p^{2} + \mu^{2})^{-1/2} \nu 
(p^{2} + \mu^{2})^{-1/2}$ 
be defined by its kernel
\begin{displaymath}
    \widetilde{\cL}_{\mu}[\nu](x,y) \defeq
    2 \mu 
    \int \frac{1}{\sqrt{p^{2} +\mu^{2}}}(x,\zeta) 
    \frac{1}{\sqrt{p^{2} +\mu^{2}}}(\zeta,y)\,\nu(d\zeta).
\end{displaymath}
Since the spectrum of an operator of the form $AA^{\dagger}$ is the 
same as that of $A^{\dagger} A$ except at zero we conclude for 
$\mu\geq 0$ 
\begin{displaymath}
	\lambda_{1}(\cL_{\mu}[\nu_{\varepsilon}]) = 
	\lambda_{1}(\widetilde{\cL}_{\mu}[\nu_{\varepsilon}]) 
	\underset{\varepsilon\to 0}{\longrightarrow}
	\lambda_{1}(\widetilde{\cL}_{\mu}[\nu]) 
\end{displaymath}
since $\lambda_{1}(\cL_{\mu}[\nu_{\varepsilon}]) > 0$ and the 
operators $\widetilde{\cL}_{\mu}[\nu_{\varepsilon}]$ converge to 
$\widetilde{\cL}_{\mu}[\nu]$ in Hilbert--Schmidt norm as 
$\varepsilon \to 0$ . Thus the equivalent of (\ref{eq:finalbound}) 
in the measure case is given by
\begin{equation}
	2\sum_{i\in \nz} \sqrt{\smash[b]{E_{i}}}
	\leq
	\nu(\rz) + 
	\lambda_{1}(\widetilde{\cL}_{\sqrt{E_{1}}}[\nu]) - 
	\lambda_{1}(\widetilde{\cL}_{\sqrt{E_{2}}}[\nu])
	\label{eq:mainmeasurebound}
\end{equation}
By the Perron--Frobenius theorem for quadratic forms we know that 
the lowest negative eigenvalue $-E_{1}$ of $p^{2} -\nu$ is simple, 
ie.\ $E_{1} > E_{2}$. So (\ref{eq:measurebound}) will follow from 
(\ref{eq:mainmeasurebound}) once we prove that 
$0\leq \mu\mapsto \lambda_{1}(\widetilde{\cL}_{\mu}[\nu])$ is 
(strictly) monotone decreasing. 
The operator $\widetilde{\cL}_{\mu}[\nu]$ is given by a strictly 
positive integral kernel and hence the eigenvector $\phi_{\mu}$ 
corresponding to the largest eigenvalue is strictly positive. 
Rewriting   
$\widetilde{\cL}_{\mu}[\nu] \phi_{\mu} = 
\lambda_{1}(\widetilde{\cL}_{\mu}[\nu]) \phi_{\mu}$ with 
$\psi_{\mu} = (p^{2} +\mu^{2})^{1/2} \phi_{\mu} > 0$ we get 
$2\mu (p^{2} \!+\!\mu^{2})^{-1} \nu \psi_{\mu} = 
\lambda_{1}(\widetilde{\cL}_{\mu}[\nu])\psi_{\mu}$. 
Consequently for $0\leq \mu_{1},\mu_{2}$
\begin{displaymath}
    \lambda_{1}(\widetilde{\cL}_{\mu_{1}}[\nu]) 
    \langle \psi_{\mu_{2}},\nu \psi_{\mu_{1}}\rangle
    =
    2\mu_{1} 
    \langle \psi_{\mu_{2}}, \nu\,
    \frac{1}{p^{2} +\mu_{1}^{2}} \,\nu
    \psi_{\mu_{1}}\rangle
\end{displaymath}
and similarly for $\lambda_{1}(\widetilde{\cL}_{\mu_{2}}[\nu])$ with 
$\mu_{1}$ and $\mu_{2}$ interchanged. As in the end of the proof 
of Lemma \ref{MonotonicityLemma} we can substract these equations 
and interchange the integration variables to arrive at  
\begin{eqnarray*}
    \lefteqn{\lambda_{1}(\widetilde{\cL}_{\mu_{1}}[\nu])  - 
    \lambda_{1}(\widetilde{\cL}_{\mu_{2}}[\nu]) } \\ 
    & = & 
    \frac{1}{\langle \psi_{\mu_{1}},\nu \psi_{\mu_{2}} \rangle} 
    \iint \nu(dx) \nu(dy) \psi_{\mu_{1}}(x)\psi_{\mu_{2}}(y) 
    \left[\e^{-\mu_{1}|x-y| } - 
      \e^{-\mu_{2}|x-y| }\right]\\
    & < & 0 \quad \text{for~} 0\leq \mu_{2} < \mu_{1}
\end{eqnarray*}
if $\nu$ is not concentrated at one point.
\end{proof}

\textbf{Acknowledgment:} D.H.\ and L.T.\ would like to thank 
  the physics department of Princeton university for its warm 
  hospitality and we thank \mbox{Wolfgang} Spitzer for discussion.
  --- The authors also thank the following 
  organizations for their support: Deutsche Forschungsgemeinschaft, 
  grant Hu 773/1-1 (DH), and the U.S. National Science Foundation, 
  grant PHY95-13072 A02 (EHL), and  grant DMS 9801329 (LET).


\begin{thebibliography}{10}
\small{

\bibitem{AizenmanLieb} 
   M.~Aizenmann and E.~H.~Lieb: 
   \textit{On semi-classical bounds for eigenvalues of
   Schr\"odinger operators.} 
   Phys.~Lett.~\textbf{66A} (1978), 427-429.

\bibitem{Birman} 
   M.~S.~Birman: 
   \textit{The spectrum of singular boundary problems.}
   Mat. Sb. \textbf{55} No.2 (1961), 125-174, translated in 
   Amer.~Math.~Soc.~Trans.\ (2), \textbf{53} (1966), 23-80.

\bibitem{Conlon}
   J.~G.~Conlon:
   \textit{A new proof of the {C}wikel-{L}ieb-{R}osenbljum 
   bound.}
   Rocky Mountain J.~Math., \textbf{15}, no.1 (1985), 117--122.

\bibitem{Cwikel} 
   M.~Cwikel: 
   \textit{Weak type estimates for singular values and the number 
   of bound states of Schr\"odinger operators.} 
   Trans.~AMS, \textbf{224} (1977), 93-100. 

\bibitem{LandauLifshitz} 
   L.~D.~Landau and E.~M.~Lifshitz: 
   \textit{Quantum Mechanics. Non-relativistic theory.} 
   Volume \textbf{3} of Course of Theoretical Physics,
   Pergamon Press (1958) 

\bibitem{LiYau} 
   P.~Li and S.-T.~Yau: 
   \textit{On the Schr\"odinger equation and the eigenvalue problem.} 
   Comm.\ Math.\ Phys., \textbf{88} (1983), 309-318.

\bibitem{Lieb} E.~H.~Lieb:
   \textit{The number of bound states of one body Schr\"odinger operators 
   and the Weyl problem.}
   Bull.\ Amer.\ Math.\ Soc., \textbf{82} (1976), 751-753.
   See also Proc.\ A.M.S.\ Symp.\ Pure Math. \textbf{36} (1980), 
   241-252. 

\bibitem{LiebLoss} E.~H.~Lieb and M.\ Loss:
   \textit{Analysis.} 
   Graduate Studies in Mathematics \textbf{14}, American Mathematical 
   Society 1997.

\bibitem{LiebThirring1975} 
   E.~H.~Lieb and W.~Thirring:
   \textit{Bound for the kinetic energy of fermions which proves 
   the stability of matter.}
   Phys.\ Rev.\ Lett., \textbf{35} (1975), 687-689. Errata 
   \textbf{35} (1975), 1116.

\bibitem{LiebThirring1976} 
   E.~H.~Lieb and W.~Thirring: 
   \textit{Inequalities for the moments of the eigenvalues of
   the Schr\"odinger Hamiltonian and their relation to Sobolev 
   inequalities.} 
   Studies in Math.\ Phys., Essays in Honor of Valentine Bargmann, 
   Princeton (1976), 
 
\bibitem{Rozenbljum} 
   G.~V.~Rozenbljum: 
   \textit{Distribution of the discrete spectrum of singular 
   differential operators.}  
   Dokl.\ AN SSSR, \textbf{202},   \textbf{N 5}  1012-1015 (1972),  
   Izv.\ VUZov, Matematika, \textbf{N.1}(1976), 75-86.

\bibitem{ReedSimonIV} 
    M.~Reed and B.~Simon: 
   \textit{Methods of modern mathematical physics IV: Analysis of 
   operators.}  
   Academic Press, New York 1978.

\bibitem{Schwinger} 
    J.~ Schwinger: 
   \textit{On the bound states of a given potential.}  
   Proc.\ Nat.\ Acad.\ Sci.\ U.S.A.\ \textbf{47}, (1961), 122--129. 


\bibitem{Simon1971} 
    B.~Simon: 
   \textit{Quantum mechanics for Hamiltonians defined as quadratic 
   forms.}  
   Princeton Series in Physics, Princeton University press, 
   New Jersey, 1971.

\bibitem{Simon1976} 
    B.~Simon: 
   \textit{The bound state of weakly coupled Schrödinger operators 
   in one and two dimensions.}  
   Ann.\ Physics \textbf{97}, no.\ 2, (1976), 279--288. 

\bibitem{Thirring} 
   W.~Thirring: 
   \textit{A course in mathematical physics. Vol. 3. Quantum mechanics 
   of atoms and molecules.} 
   Translated from the German by Evans M. Harrell. 
   Lecture Notes in Physics, \textbf{141}. Springer-Verlag, 
   New York-Vienna, 1981.

\bibitem{Weidl} 
   T.~Weidl: 
   \textit{On the Lieb-Thirring constants $L_{\gamma,1}$ for $\gamma\geq 
   1/2.$}   
   Comm.\ Math.\ Phys., \textbf{178}, no.\ 1, (1996), 135--146. 

}
\end{thebibliography}
\end{document}